\documentclass[a4paper,11pt]{article}
\usepackage{pos}

\usepackage{slashed}
\usepackage{changes}
\usepackage{tikz}
\usetikzlibrary{arrows,shapes}
\usetikzlibrary{shapes.multipart}
\usepackage[compat=1.1.0]{tikz-feynman}
\usepackage{contour}
\usepackage{feynmp}
\usepackage{tikzpagenodes}
\usepackage{subcaption}
\usepackage{float}
\usepackage{xspace}
\newcounter{comment}

\newcommand{\CMcolor}{violet} 
\newcommand{\JMcolor}{red}
\newcommand{\ACcolor}{orange}
\newcommand{\BPcolor}{blue}

{\refstepcounter{comment}%
\begin{quote}
\color{\CMcolor}{#1~ \normalsize \textbf{\underline{Comment} $\sharp$\thecomment~by~CM:~}}}%
{\end{quote}}

{\refstepcounter{comment}%
\begin{quote}
\color{\JMcolor}{#1~ \normalsize \textbf{\underline{Comment} $\sharp$\thecomment~by~JM:~}}}%
{\end{quote}}

{\refstepcounter{comment}%
\begin{quote}
\color{\ACcolor}{#1~ \normalsize \textbf{\underline{Comment} $\sharp$\thecomment~by~AC:~}}}%
{\end{quote}}

{\refstepcounter{comment}%
\begin{quote}
\color{\BPcolor}{#1~ \normalsize \textbf{\underline{Comment} $\sharp$\thecomment~by~BP:~}}}%
{\end{quote}}

\definechangesauthor[color=\CMcolor]{CM}
\definechangesauthor[color=\JMcolor]{JM}
\definechangesauthor[color=\BPcolor]{BP}
\definechangesauthor[color=\ACcolor]{AC}

\title{Backward DVCS on the pion in Sullivan processes}

\author*[a]{Abigail Castro}
\author[a]{Cedric Mezrag}
\author[a]{Jose M. Morgado Chávez}
\author[b]{Bernard Pire}

\affiliation[a]{ IRFU, CEA, Universit\'e Paris-Saclay,
  91191 Gif Sur Yvette, France}

\affiliation[b]{CPHT, CNRS, Ecole polytechnique, Institut Polytechnique de Paris, 
91128 Palaiseau, France}

\emailAdd{abigail.rodriguescastro@cea.fr}
\emailAdd{cedric.mezrag@cea.fr}
\emailAdd{jose-manuel.morgadochavez@cea.fr}
\emailAdd{bernard.pire@polytechnique.edu}
\abstract{The purpose of this work is to do a systematic feasibility study of measuring in backward region deeply virtual Compton scattering on the pion in Sullivan processes in the framework of collinear QCD factorization where pion to photon transition distribution amplitudes (TDAs) describe the photon content of the $\pi$ meson. Our approach employs TDAs based on the overlap of light front wave functions, using a previously developed pion light-front wave function and deriving a consistent model for the light front wave functions of the photon.
This work is expected to lead us to an estimate of the cross-sections that could be measured in the future U.S. and China's electron-ion colliders. It will also provide a comparison with the forward Sullivan DVCS case, which gives access to pion GPDs and for which a strong signal is expected.}

\FullConference{31st International Workshop on Deep Inelastic Scattering (DIS2024)\\
 8–12 April 2024\\
Grenoble, France\\}

\begin{document}
\maketitle
\section{Forward and backward DVCS in a Sullivan process}
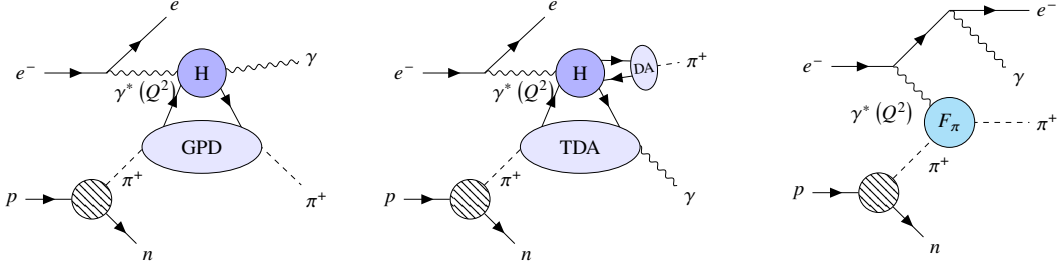
\begin{figure}[htp]
\centering
\scalebox{0.7}[0.7]{
\begin{subfigure}[H]{0.45\textwidth}
\centering
\begin{tikzpicture}
  \begin{feynman}
    \vertex (a) {\(e^{-}\)};
    \vertex [right=of a] (b);
    \vertex [above right=of b] (f1) {\(e^{-}\)};
    \vertex [below right=of b] (c);
    \vertex [right=of b] (c2);
    \vertex [blob, below left=of c] (d) {\contour{white}{}};
    \vertex [left =of d] (e) {\(p\)};
    \vertex [below right= of d] (f) {\(n\)};
    \vertex [right =of c] (g);
    \vertex [above right =of g] (h) {\(\gamma\)};
    \vertex [below right =of g] (i) {\(\pi^{+}\)};
    \vertex (m) at (3.3,-1.4);
    \vertex (n) at (3.3,-0);
    \vertex (g2) at (3.5,0) ;
    \vertex (pi) at (2, -2) {\(\pi^{+}\)};
    
    \diagram* {
      (a) -- [fermion] (b) -- [fermion] (f1),
      (b) -- [boson, edge label'=\(\gamma^{*}\left(Q^{2}\right)\)] (c2),
      (g2) -- [boson] (h),
      (g) -- [scalar] (i),
      (d) -- [scalar] (c),
      (e) -- [fermion] (d),
      (d) -- [fermion] (f),
      (c) -- [fermion] (c2),
      (g2) -- [fermion] (g),
    };  
  \end{feynman}  
   \node[draw,fill=blue!30,circle, minimum height=0.5cm, minimum width=0.5cm,align=center, text width=0.5cm]  at (n) {H};
   \node[draw,fill=blue!10,ellipse, minimum height=1cm, minimum width=2.25cm,align=center, text width=1cm] at (m) {GPD};
\end{tikzpicture}
\end{subfigure}
}
\scalebox{0.7}[0.7]{
\begin{subfigure}[H]{0.45\textwidth}
\centering
\begin{tikzpicture}
  \begin{feynman}
    \vertex (a) {\(e^{-}\)};                                    
    \vertex [right=of a] (b);                                   
    \vertex [above right=of b] (f1) {\(e^{-}\)};                
    \vertex [below right=of b] (c);                             
    \vertex [right=of b] (c2);                                  
    \vertex [blob, below left=of c] (d) {\contour{white}{}};    
    \vertex [left =of d] (e) {\(p\)};                           
    \vertex [below right= of d] (f) {\(n\)};                    
    \vertex [right =of c] (g);                                  
    \vertex [below right =of g] (i) {\(\gamma\)};               

    \vertex (A) at (3.3-0.25*0.2/1.8,0.25); 
    \vertex (B) at (3.3+0.3*0.2/1.8,-0.2) {\(\pi^{+}\)};

    \vertex (C) at (3.3+1.2-0.2*0.1/1.8,0.1+0.2*1.1/1.8);
    \vertex (D) at (3.3+1.2+0.2*0.1/1.8,0.1-0.2*1.1/1.8);

    \vertex (Pi1) at (4.5,0.1);
    \vertex (Pi2) at (3.3+2.2,0.25) {\(\pi^{+}\)};
    
    \vertex (m) at (3.3,-1.4);                                  
    \vertex (n) at (3.3,-0);
    \vertex (g2) at (3.5,0) ;
    
    \vertex (pi) at (2, -2) {\(\pi^{+}\)};
    
    \diagram* {
      (a) -- [fermion] (b) -- [fermion] (f1),
      (b) -- [boson, edge label'=\(\gamma^{*}\left(Q^{2}\right)\)] (c2),
      (g) -- [boson] (i),
      (d) -- [scalar] (c),
      (e) -- [fermion] (d),
      (d) -- [fermion] (f),
      (c) -- [fermion] (c2),
      (g2) -- [fermion] (g),
      (Pi1) -- [scalar] (Pi2);
      (C) -- [anti fermion] (A),
      (B) -- [anti fermion] (D)
    };  
  \end{feynman}  
   \node[draw,fill=blue!30,circle, minimum height=0.5cm, minimum width=0.5cm,align=center, text width=0.5cm]  at (n) {H};
   \node[draw,fill=blue!10,ellipse, minimum height=1cm, minimum width=2.25cm,align=center, text width=1cm] at (m) {TDA};
   \node[draw,fill=blue!10,ellipse, minimum height=0.85cm, minimum width=0.5cm,align=center, rotate=6.54] at (Pi1) {};
    \node[rotate=6.34] at (Pi1) {\scriptsize DA};
\end{tikzpicture}
\end{subfigure}
}
\scalebox{0.7}[0.7]{
\begin{subfigure}[H]{0.45\textwidth}
\centering
\begin{tikzpicture}
  \begin{feynman}
    \vertex (a) {\(e^{-}\)};
    \vertex [right=of a] (b);
    \vertex [above right =of b] (c);
    \vertex [right =of c] (d) {\(e^{-}\)};
    \vertex [below right =of c] (g) {\(\gamma\)};
    
    \vertex [below right =of b] (e);
    
    \vertex [right =of e] (f) {\(\pi^{+}\)};
    
    \vertex [blob, below left =of e] (h) {\contour{white}{}};
    \vertex [left =of h] (i) {\(p\)};
    \vertex [below right =of h] (j) {\(n\)};
    
    \diagram* {
      (a) -- [fermion] (b) -- [fermion] (c) -- [fermion] (d),
      (b) -- [boson, edge label'=\(\gamma^{*}\left(Q^{2}\right)\)] (e),
      (h) -- [scalar, edge label'=\(\pi^{+}\)] (e),
	   (i) -- [fermion] (h),
	   (h) -- [fermion] (j),
	   (e) -- [scalar] (f),
	   (c) -- [boson] (g),
    };
    
  \end{feynman}
  
  \node[draw,fill=cyan!30,circle, minimum height=0.5cm, minimum width=0.5cm,align=center, text width=0.5cm] at (e) {\(F_{\pi}\)};
      
\end{tikzpicture}
\end{subfigure}
}
\caption{\textsc{Left panel}: forward DVCS; \textsc{Central panel}: backward DVCS; \textsc{Right panel}: the Bethe-Heitler process, in the Sullivan reaction  $e^-(l) p(p)  \to  e^-(l') \gamma(q') \pi^+(p'_\pi) n(p')$.  }
\label{fig:kin}
\end{figure}

Hard exclusive reactions are the golden way to perform quark and gluon tomography of hadrons. The tomography of mesons is a difficult task since there is no meson target. To circumvent this difficulty, the Sullivan processes \cite{sullivan} consider quasi-real $\pi-$mesons emitted by a nucleon target. Near forward deeply virtual Compton scattering in a Sullivan process (see Fig.~\ref{fig:kin} -- left panel) has indeed been proposed \cite{Amrath:2008vx} to extract $ \pi -$meson leading twist generalized parton distributions (GPDs) and feasibility studies performed \cite{Chavez:2021llq,Chavez:2021koz}.
Backward processes have recently been the subject of a renewed interest \cite{Gayoso:2021rzj}, in particular in the context of a factorized description of their amplitudes in terms of transition distribution amplitudes \cite{Pire:2004ie, Pire:2021hbl} which generalize the notion of GPDs.
We thus consider the reactions
\begin{equation}
  \label{proc}
e(l) + p(p) \to e(l') + \gamma(q') + \pi^+(p'_\pi) + n(p') \,,
\end{equation}
in the near backward region where $-u_\pi = -(q-p'_\pi)^2$ is small,
with $q = l-l' \,$ and $ 
p_\pi = p-p'\,$, the virtual photon and  $\pi$ meson momenta respectively, we define 
the energy fractions $x_B = \frac{Q^2}{2(p-p')\cdot q}$, ~$\xi=\frac{x_B}{2-x_B}$
.

In this context, we identify two contributions to the amplitude of reaction Eq.~\eqref{proc}: a strong process, backward deeply virtual Compton scattering (bDVCS, Fig.~\ref{fig:kin} -- central panel); and a purely electromagnetic one, the Bethe-Heitler process  (see right pannel of Fig.\ref{fig:kin}). The latter has a negligible amplitude at small values of $-u_\pi$ and we can ignore it in our analysis. We thus focus, solely, on the bDVCS contribution.

\section{Backward DVCS amplitude}

From now on, we assume that the pion source part of the Sullivan process can be factorised (see \cite{Amrath:2008vx}) and we focus on the bDVCS part of the diagram. 
The proof of factorization of the bDVCS amplitude as a convolution of a short distance coefficient function ($C_{F}$), a meson distribution amplitude ($\Phi^\pi$) and a $\pi \to \gamma$ TDA ($A^{\pi\gamma}$) follows the line of the factorization proof of meson forward deep electroproduction \cite{Collins:1996fb} on a nucleon, with the nucleon GPD replaced by the TDA. The leading twist QCD amplitude $\mathcal{A}_L$ for the process $\gamma^*_L \pi^{+} \to \pi^{+} \gamma$ thus reads 
 \begin{equation}
  \label{ampDVCS}
  \mathcal{A}_L^{\pi^+} (\xi,u,Q^2) = \frac{16\pi\alpha_s e}{9Q} \int dx dz C^{ud}_{F}(x,z,\xi) \Phi^{\pi^+}(z)A^{\pi^+\gamma} (x, \xi,u) \,,
\end{equation}
where  $C_F$ reads at leading order \cite{Diehl:2003ny}
\begin{equation}
  \label{CF}
C^{qq'}_{F}(x,z,\xi) = \frac{1}{1-z}\frac{e_q}{\xi-x-i\varepsilon} - \frac{1}{z}\frac{e_{q'}}{\xi+x-i\varepsilon}\,, \quad q \neq q'.
\end{equation}
Since the pion DA is symmetric in $z \to (1-z)$, the $z-$integration factorizes in a prefactor $\int dz \Phi^\pi(z)/z$. 

\section{Model for the \texorpdfstring{$\pi \to \gamma$}{pi-to-gamma} TDAs}
There are four leading twist $\pi \to \gamma$ TDAs: one vector, one axial and two transversity. In our process, only the axial quark TDA $A^\pi$ contributes. It is defined as 
\begin{equation} \label{defH}
\frac{e}{f_{\pi}}\epsilon\cdot\Delta  A^{\pi^+ \gamma} =\frac{1}{2}\int\frac{dz^{-}}{2\pi}e^{ixP^{+}z^{-}}\left.\left\langle\gamma,P+\frac{\Delta}{2}\right|\overline{\psi}_{q'}\left(-\frac{z}{2}\right)\gamma^{+}\gamma_{5}\psi_{q}\left(\frac{z}{2}\right)\left|\pi^{+},P-\frac{\Delta}{2}\right\rangle\right|_{z^{+}=z^{\perp}_{i}=0}\,.
\end{equation}
 where $\epsilon$ is the outgoing photon polarisation vector and $f_\pi$ the pion decay constant. Few models for the $\pi \to \gamma$ TDAs already exist \cite{Tiburzi:2005nj, Lansberg:2006fv,Courtoy:2007vy}. The starting point for ours is the lowest Fock state description of a $\pi^+$ meson wave function :
\begin{eqnarray}
    \label{eq:PionKet0}
    |\pi^+,\uparrow \downarrow \rangle &= & \int \frac{\textrm{d}k_\perp}{16\pi^3} \frac{x}{\sqrt{x(1-x)}} \psi_{\uparrow \downarrow} \left[b_{u,\uparrow}^\dagger(x,k_\perp) d_{d,\downarrow}^\dagger(1-x,-k_\perp)\right.\nonumber \\
    && -\left.  b_{u,\downarrow}^\dagger(x,k_\perp) d_{d,\uparrow}^\dagger(1-x,-k_\perp) \right] |0 \rangle
\end{eqnarray}
with the Light-Front wave functions \cite{Chouika:2017rzs} (LFWF) given as:
\begin{align}
    \label{eq:PionLFWF0}
    \psi^\pi_{\uparrow\downarrow}(x,k_\perp) 
    = 8 \sqrt{15} \pi \frac{M^3}{(k_\perp^2 + M^2)^2} x(1-x), 
\end{align}
where $M$ is a mass scale fitted to $M=318~ \textrm{MeV}$. The pion presents a second independent LFWF, $\psi^\pi_{\uparrow\uparrow}$ associated with the Fock state $ |\pi^+,\uparrow \uparrow\rangle $, whose computation to the contribution to the TDA is ongoing. For the photon case, the Fock state decomposition was employed as presented in \cite{Diehl:2003ny} to obtain the photon states, and the Light-Front wave functions were derived based on the methodologies outlined in \cite{Dorokhov:2006qm, Ball:2002ps} (see \cite{Shi:2023jyk} for an alternative discussion).

In such a two-body approach, the TDA can be further decomposed into flavour contributions, labelling the quark flavour involved in the formation of the outgoing photon,
\begin{align}
    \label{eq:FLavourDecomposition}
    A^{\pi^+ \gamma} = e \left(e_u A^{\pi^+ \gamma}_{u}+e_{d} A^{\pi^+ \gamma}_{d}\right)\,.
\end{align}

Using the overlap method developed for GPDs \cite{Diehl:2000xz}, we obtain the TDA in terms of these LFWFs\footnote{Applying this method and using our photon LFWF allows us to recover the anomalous GPD of the photon \cite{Friot:2006mm}.}
in the DGLAP region  $x\geq|\xi|$  in closed form:
\begin{equation}\label{eqcov:TDAmodel}
\left.A^{\pi^{+}\gamma}_{q}\left(x,\xi,t\right)\right|_{x\geq|\xi|}=\left.A^{\pi^{+}\gamma}_{q(\uparrow\downarrow)}\left(x,\xi,t\right)\right|_{x\geq|\xi|}+\left.A^{\pi^{+}\gamma}_{q(\uparrow\uparrow)}\left(x,\xi,t\right)\right|_{x\geq|\xi|}\,,
\end{equation}
where $(\uparrow\downarrow)$ and $(\uparrow\uparrow)$ labels the quark helicity projections, and thus the different LFWFs contributions to the TDA.
\begin{figure}[!ht]
\centering
\begin{subfigure}[h]{0.45\textwidth}
\includegraphics[scale=0.75]{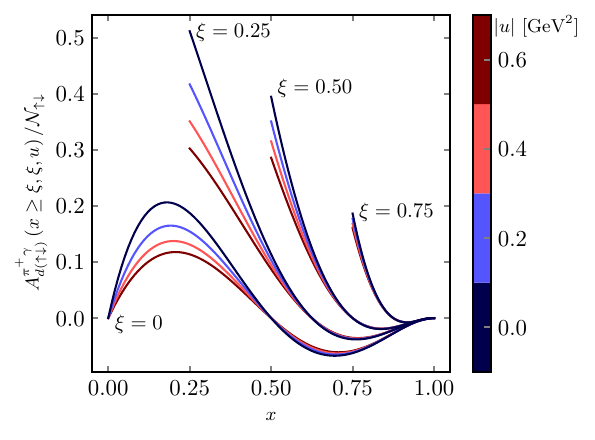}    \end{subfigure}
\hfill
\begin{subfigure}[h]{0.45\textwidth}
\includegraphics[scale=0.75]{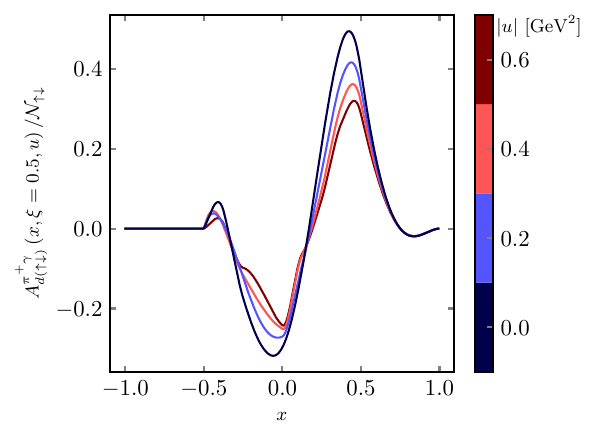}    
\end{subfigure}
\caption{\label{figcov:ddt0} TDA $A^{\pi^+\gamma}_{d(\uparrow \downarrow)}$ constructed in this study, evaluated at selected $u$-values (see legend). \textsc{Left panel}: DGLAP region, Eq.~\eqref{eqcov:TDAanti}, obtained from the overlap of pion and photon light-front wave functions. The curves, ordered from left to right, correspond to $\xi \in {0,0.25,0.5,0.75}$ . For each $\xi$ value, the TDA is calculated at four fixed $u$ values $(0,0.2,0.4,0.6)$  represented by the following colors: dark blue, light blue, red, and dark red, respectively. \textsc{Right panel}: Full kinematic domain, including the ERBL region obtained through the covariant extension strategy, represented at $\xi=0.5$.}
\end{figure}

As an illustration, consider the contribution to the $\pi\rightarrow\gamma$ TDA generated by the $d\overline{d}$ Fock-space-expansion of the photon state
(i.e. the $u$ quark of the $\pi^+$ enters the hard kernel):
\begin{equation}\label{eqcov:TDAanti}
\left.A^{\pi^{+}\gamma}_{d(\uparrow\downarrow)}\left(x,\xi,u\right)\right|_{x\geq |\xi|}=\mathcal{N}_{\uparrow\downarrow}\frac{\left(1-x\right)^{2}\left(x+\xi\right)}{\left(1-\xi^{2}\right)^{2}\left(1+\xi\right)}\left[\left(\xi-x\right)+\left(1-x\right)\right]\frac{\tau\left(2\tau+1\right)-\sqrt{\frac{\tau}{\tau+1}}\tanh^{-1}\left(\sqrt{\frac{\tau}{\tau+1}}\right)}{\tau^{2}\left(1+\tau\right)}.
\end{equation}
where $\tau= -\frac{(1-x)^2}{(1-\xi^2)} \frac{u}{4M^2}$.
According to the covariant extension strategy \cite{Chouika:2017dhe, Chavez:2021llq, DallOlio:2024vjv}, the knowledge of TDAs within the DGLAP region uniquely specifies their ERBL domain. In a nutshell: as GPDs, $\pi \to \gamma$ TDAs benefit from a representation as the Radon transform of double distributions\footnote{Importantly, TDAs being flavor non-singlet objects, no D-term--like ambiguity arises, in contrast to the GPD case.}. Provided that the solution to the inverse Radon transform problem exists and is unique when TDAs are known only on the DGLAP region \cite{Chouika:2017dhe}, the associated double distribution can be found and employed afterwards to reconstruct the ERBL domain \cite{Chavez:2021llq, DallOlio:2024vjv}. For the case above, Eq.~\eqref{eqcov:TDAanti}, in the $u\rightarrow 0$, the double distribution, $h$, is found to be a polynomial in the kinematic variables $\left(\beta,\alpha\right)$
\begin{equation}
h^{\pi^{+}\gamma}_{d(\uparrow\downarrow)}\left(\beta,\alpha,0\right) = -\mathcal{N}_{\uparrow\downarrow}\left[\frac{1}{3}-\frac{10}{3}\alpha-\alpha^{2}+4\alpha^{3}-\frac{10}{3}\beta+6\alpha\beta+4\alpha^{2}\beta+7\beta^{2}-4\alpha\beta^{2}-4\beta^{3}\right],
\end{equation}
which yields
\begin{equation}
\begin{array}{rcl}
\displaystyle \left.A^{\pi^{+}\gamma}_{d(\uparrow\downarrow)}\left(x,\xi,0\right)\right|_{x\leq|\xi|}&\displaystyle = & \displaystyle \frac{\mathcal{N}_{\uparrow\downarrow}}{3\xi^{4}\left(1+\xi\right)^{3}}\left[x^{2}\xi^{2}\left(5+\xi\left(20+3\xi\right)\right)-x^{4}\left(3+\xi\left(10+11\xi\right)\right)\right.\\
&&\displaystyle \\
&&\left.-2\xi^{4}\left(1+\xi\right)-x\xi^{3}\left(1-\xi\left(8+5\xi\right)\right)+x^{3}\xi\left(1-13\xi^{2}\right)\right].\\
\end{array}
\end{equation}

In the general case where $u\neq 0$,  we explore the numerical procedure for the solution of the inverse Radon transform problem described in \cite{Chavez:2021llq,DallOlio:2024vjv}. This allows us to get a parametrization of the double distribution, from which we calculate the TDA. Our results are shown on Fig. \ref{figcov:ddt0}. Note that contrary to the GPD case, no symmetry in $\xi$ can help improve the numerical computations.

In our model, the $[-x,-\xi]$ region is also contributing to the amplitude thanks to the symmetry relation between the $A^{\pi^+\gamma}_{u(\uparrow \downarrow)}$ contribution and the $A^{\pi^+\gamma}_{d(\uparrow \downarrow)}$ one:
\begin{align}
    A^{\pi^+\gamma}_{u(\uparrow \downarrow)}(x,\xi,u) = A^{\pi^+\gamma}_{d(\uparrow \downarrow)}(-x, \xi,u),
\end{align}
\section{Conclusion}

We have presented here our preliminary result in our attempt to evaluate the measurability of Sullivan Backward DVCS at existing and future facilities. As expected, the formalism developed in the case of forward Sullivan DVCS can be adapted to the backward case, with though a few additions and complications. Yet, we demonstrated how the amplitude can be assessed in a simple LFWFs model. Before computing the amplitude itself, one needs to take into account of the second contribution to the TDA, with aligned quark helicities. We foresee no additional difficulties, and we expect to obtain the backward DVCS amplitude soon, after evolving the TDA to scales relevant for current and future experimental facilities. Refinement can be envisioned, such as NLO corrections, and other processes, like TCS, could be addressed. Last but not least, replacing the produced $\pi^+$ meson by a longitudinally polarized $\rho^+$ meson will test the vector $\pi \to \gamma$ TDA. 

\acknowledgments
We acknowledge useful discussions with Maxime Defurne, Kirill Semenov-Tian-Shansky and Lech Szymanowski.
This research was funded in part by l’Agence Nationale de la Recherche (ANR), project ANR-23-CE31-0019 and  by the Coordenação de Aperfeiçoamento de Pessoal de Nível
Superior - Brasil (CAPES) – Finance Code 001. For the purpose of open access, the authors have applied a CC-BY public copyright licence to any Author Accepted Manuscript (AAM) version arising from this submission.

\bibliographystyle{JHEP}
\bibliography{DIS}

\providecommand{\noopsort}[1]{}\providecommand{\singleletter}[1]{#1}%

\providecommand{\href}[2]{#2}\begingroup\raggedright\begin{thebibliography}{10}

\bibitem{sullivan}
J.D.~Sullivan, \emph{{One pion exchange and deep inelastic electron - nucleon scattering}}, \href{https://doi.org/10.1103/PhysRevD.5.1732}{\emph{Phys. Rev. D} {\bfseries 5} (1972) 1732}.

\bibitem{Amrath:2008vx}
D.~Amrath, M.~Diehl and J.-P.~Lansberg, \emph{{Deeply virtual Compton scattering on a virtual pion target}}, \href{https://doi.org/10.1140/epjc/s10052-008-0769-1}{\emph{Eur. Phys. J. C} {\bfseries 58} (2008) 179} [\href{https://arxiv.org/abs/0807.4474}{{\ttfamily 0807.4474}}].

\bibitem{Chavez:2021llq}
J.M.M.~Chavez, V.~Bertone, F.~De~Soto~Borrero, M.~Defurne, C.~Mezrag, H.~Moutarde et~al., \emph{{Pion generalized parton distributions: A path toward phenomenology}}, \href{https://doi.org/10.1103/PhysRevD.105.094012}{\emph{Phys. Rev. D} {\bfseries 105} (2022) 094012} [\href{https://arxiv.org/abs/2110.06052}{{\ttfamily 2110.06052}}].

\bibitem{Chavez:2021koz}
J.M.M.~Ch\'avez, V.~Bertone, F.~De~Soto~Borrero, M.~Defurne, C.~Mezrag, H.~Moutarde et~al., \emph{{Accessing the Pion 3D Structure at US and China Electron-Ion Colliders}}, \href{https://doi.org/10.1103/PhysRevLett.128.202501}{\emph{Phys. Rev. Lett.} {\bfseries 128} (2022) 202501} [\href{https://arxiv.org/abs/2110.09462}{{\ttfamily 2110.09462}}].

\bibitem{Gayoso:2021rzj}
C.A.~Gayoso et~al., \emph{{Progress and opportunities in backward angle (u-channel) physics}}, \href{https://doi.org/10.1140/epja/s10050-021-00625-2}{\emph{Eur. Phys. J. A} {\bfseries 57} (2021) 342} [\href{https://arxiv.org/abs/2107.06748}{{\ttfamily 2107.06748}}].

\bibitem{Pire:2004ie}
B.~Pire and L.~Szymanowski, \emph{{Hadron annihilation into two photons and backward VCS in the scaling regime of QCD}}, \href{https://doi.org/10.1103/PhysRevD.71.111501}{\emph{Phys. Rev. D} {\bfseries 71} (2005) 111501} [\href{https://arxiv.org/abs/hep-ph/0411387}{{\ttfamily hep-ph/0411387}}].

\bibitem{Pire:2021hbl}
B.~Pire, K.~Semenov-Tian-Shansky and L.~Szymanowski, \emph{{Transition distribution amplitudes and hard exclusive reactions with baryon number transfer}}, \href{https://doi.org/10.1016/j.physrep.2021.09.002}{\emph{Phys. Rept.} {\bfseries 940} (2021) 1} [\href{https://arxiv.org/abs/2103.01079}{{\ttfamily 2103.01079}}].

\bibitem{Collins:1996fb}
J.C.~Collins, L.~Frankfurt and M.~Strikman, \emph{{Factorization for hard exclusive electroproduction of mesons in QCD}}, \href{https://doi.org/10.1103/PhysRevD.56.2982}{\emph{Phys. Rev. D} {\bfseries 56} (1997) 2982} [\href{https://arxiv.org/abs/hep-ph/9611433}{{\ttfamily hep-ph/9611433}}].

\bibitem{Diehl:2003ny}
M.~Diehl, \emph{{Generalized parton distributions}}, \href{https://doi.org/10.1016/j.physrep.2003.08.002}{\emph{Phys. Rept.} {\bfseries 388} (2003) 41} [\href{https://arxiv.org/abs/hep-ph/0307382}{{\ttfamily hep-ph/0307382}}].

\bibitem{Tiburzi:2005nj}
B.C.~Tiburzi, \emph{{Estimates for pion-photon transition distributions}}, \href{https://doi.org/10.1103/PhysRevD.72.094001}{\emph{Phys. Rev. D} {\bfseries 72} (2005) 094001} [\href{https://arxiv.org/abs/hep-ph/0508112}{{\ttfamily hep-ph/0508112}}].

\bibitem{Lansberg:2006fv}
J.P.~Lansberg, B.~Pire and L.~Szymanowski, \emph{{Exclusive meson pair production in gamma* gamma scattering at small momentum transfer}}, \href{https://doi.org/10.1103/PhysRevD.73.074014}{\emph{Phys. Rev. D} {\bfseries 73} (2006) 074014} [\href{https://arxiv.org/abs/hep-ph/0602195}{{\ttfamily hep-ph/0602195}}].

\bibitem{Courtoy:2007vy}
A.~Courtoy and S.~Noguera, \emph{{The Pion-photon transition distribution amplitudes in the Nambu-Jona Lasinio model}}, \href{https://doi.org/10.1103/PhysRevD.76.094026}{\emph{Phys. Rev. D} {\bfseries 76} (2007) 094026} [\href{https://arxiv.org/abs/0707.3366}{{\ttfamily 0707.3366}}].

\bibitem{Chouika:2017rzs}
N.~Chouika, C.~Mezrag, H.~Moutarde and J.~Rodr\'\i{}guez-Quintero, \emph{{A Nakanishi-based model illustrating the covariant extension of the pion GPD overlap representation and its ambiguities}}, \href{https://doi.org/10.1016/j.physletb.2018.02.070}{\emph{Phys. Lett. B} {\bfseries 780} (2018) 287} [\href{https://arxiv.org/abs/1711.11548}{{\ttfamily 1711.11548}}].

\bibitem{Dorokhov:2006qm}
A.E.~Dorokhov, W.~Broniowski and E.~Ruiz~Arriola, \emph{{Photon distribution amplitudes and light-cone wave functions in chiral quark models}}, \href{https://doi.org/10.1103/PhysRevD.74.054023}{\emph{Phys. Rev. D} {\bfseries 74} (2006) 054023} [\href{https://arxiv.org/abs/hep-ph/0607171}{{\ttfamily hep-ph/0607171}}].

\bibitem{Ball:2002ps}
P.~Ball, V.M.~Braun and N.~Kivel, \emph{{Photon distribution amplitudes in QCD}}, \href{https://doi.org/10.1016/S0550-3213(02)01017-9}{\emph{Nucl. Phys. B} {\bfseries 649} (2003) 263} [\href{https://arxiv.org/abs/hep-ph/0207307}{{\ttfamily hep-ph/0207307}}].

\bibitem{Shi:2023jyk}
C.~Shi, Z.~Yang, X.~Chen, W.~Jia, C.~Luo and W.~Xiang, \emph{{Nonperturbative photon qq\textasciimacron{} light front wave functions from a contact interaction model}}, \href{https://doi.org/10.1103/PhysRevD.109.034020}{\emph{Phys. Rev. D} {\bfseries 109} (2024) 034020} [\href{https://arxiv.org/abs/2310.19042}{{\ttfamily 2310.19042}}].

\bibitem{Diehl:2000xz}
M.~Diehl, T.~Feldmann, R.~Jakob and P.~Kroll, \emph{{The overlap representation of skewed quark and gluon distributions}}, \href{https://doi.org/10.1016/S0550-3213(00)00684-2}{\emph{Nucl. Phys. B} {\bfseries 596} (2001) 33} [\href{https://arxiv.org/abs/hep-ph/0009255}{{\ttfamily hep-ph/0009255}}].

\bibitem{Friot:2006mm}
S.~Friot, B.~Pire and L.~Szymanowski, \emph{{Deeply virtual compton scattering on a photon and generalized parton distributions in the photon}}, \href{https://doi.org/10.1016/j.physletb.2006.12.038}{\emph{Phys. Lett. B} {\bfseries 645} (2007) 153} [\href{https://arxiv.org/abs/hep-ph/0611176}{{\ttfamily hep-ph/0611176}}].

\bibitem{Chouika:2017dhe}
N.~Chouika, C.~Mezrag, H.~Moutarde and J.~Rodr\'\i{}guez-Quintero, \emph{{Covariant Extension of the GPD overlap representation at low Fock states}}, \href{https://doi.org/10.1140/epjc/s10052-017-5465-6}{\emph{Eur. Phys. J. C} {\bfseries 77} (2017) 906} [\href{https://arxiv.org/abs/1711.05108}{{\ttfamily 1711.05108}}].

\bibitem{DallOlio:2024vjv}
P.~Dall'Olio, F.~De~Soto, C.~Mezrag, J.M.~Morgado~Ch\'avez, H.~Moutarde, J.~Rodr\'\i{}guez-Quintero et~al., \emph{{Unraveling generalized parton distributions through Lorentz symmetry and partial DGLAP knowledge}}, \href{https://doi.org/10.1103/PhysRevD.109.096013}{\emph{Phys. Rev. D} {\bfseries 109} (2024) 096013} [\href{https://arxiv.org/abs/2401.12013}{{\ttfamily 2401.12013}}].

\end{thebibliography}\endgroup
\end{document}